\documentclass[
 reprint,
 superscriptaddress,
 amsmath,amssymb,
 aps,
 pra,
]{revtex4-2}

\usepackage{graphicx}
\usepackage{dcolumn}
\usepackage{bm}

\usepackage{lipsum}
\usepackage{indentfirst}
\usepackage{verbatim}
\usepackage{xcolor}
\usepackage{ulem}
\usepackage{units}

\newcommand{\TITLE}{Shift of the Bose-Einstein condensation transition in the presence of a \\ second atomic species}

\begin{document}

\preprint{APS/123-QED}

\title{\TITLE}

\author{P.M. Gaspar}
    \altaffiliation[Present Address: Université Sorbonne Paris Nord, Laboratoire de Physique des Lasers, CNRS UMR 7538, 99 av. J.-B. Clément, F-93430 Villetaneuse, France]{}
    \affiliation{Instituto de F\'isica de S\~ao Carlos, Universidade de S\~ao Paulo, CP 369, 13560-970 S\~ao Carlos, Brazil.}
    
\author{V.S. Bagnato}
    \affiliation{Instituto de F\'isica de S\~ao Carlos, Universidade de S\~ao Paulo, CP 369, 13560-970 S\~ao Carlos, Brazil.}
    \affiliation{Texas A\&M University - BME and Physics Department, College Station, TX, USA}
	
\author{P.C.M. Castilho}
        \email[Corresponding author: ]{patricia.castilho@ifsc.usp.br}
	\affiliation{Instituto de F\'isica de S\~ao Carlos, Universidade de S\~ao Paulo, CP 369, 13560-970 S\~ao Carlos, Brazil.}
    





\date{\today}

\begin{abstract}
Atomic interactions play an important role in the properties of ultracold atomic gases. In single component bosonic systems, its effect is already present at the critical point for the Bose-Einstein condensate phase transition by shifting it to lower temperatures as a consequence of effective repulsion between the atoms. When considering atomic bosonic mixtures, interesting effects arise from the competition between intra- and interspecies interactions such as the miscible-immiscible phase transition and the particular case of self-bounded quantum droplets. In such a scenario, it is natural to expect that these interactions will also affect the critical point of each species composing the mixture. In this paper, we obtain semi-analytical expressions for the critical temperature shift of the phase transition to a Bose-Einstein condensate in the presence of a second species considering the mean-field framework. We treat differently the cases in which the second species is above or below its own critical temperature and apply the obtained relations to the case of a $^{23}$Na-$^{39}$K bosonic mixture which can be realized in current running experimental setups. Our findings can be easily extended to other atomic mixtures.
\end{abstract}


\maketitle


\section{Introduction}\label{Sec:Intro}

Over the past three decades, ultracold atom experiments have become versatile platforms to study phenomena ranging from the basis of quantum mechanics \cite{anderson1995observation,davis1995bose,bradley1995evidence} to cosmology \cite{steinhauer2016observation} and condensed matter physics~\cite{gross2017quantum}. Most of its success is related to the high level of control of the system's parameters and to the precision of their theoretical descriptions. In this sense, real ultracold atoms normally interact among themselves via a contact interaction governed by $s$-wave scattering with control parameter given by the $s$-wave scattering length, $a$. This interaction has strong impact on the system's properties as for example, its own superfluid nature which is suppressed for an ideal gas even at absolute zero as a result of Landau's criterion for superfluidity \cite{pitaevskii2016bose}. It is due to interactions that one can explore the impurity problem in the context of Bose and Fermi polarons \cite{bruderer2008self,spethmann2012dynamics}, produce ultracold gases made of molecules \cite{park2015ultracold} or explore the phase-diagram of ultracold atomic mixtures in different scenarios~\cite{ho1996binary,esry1997hartree,pu1998properties,riboli2002topology,altman2003phase,isacsson2005superfluid}.

Not only the properties of ultracold atomic systems are affected by interactions, but when considering phase transitions, their critical point will experience a shift. This has been well-known since the early days of realizing atomic Bose-Einstein condensates (BECs). For an ideal Bose gas of $N$ atoms trapped by a harmonic potential given by $V(\textbf{r}) = \frac{1}{2}m(\omega_x^2x^2 + \omega_y^2y^2 + \omega_z^2z^2)$, where $m$ is the atomic mass and $\omega_{x,y,z}$ are the angular frequencies along the $\widehat{x}$, $\widehat{y}$ and $\widehat{z}$ axis, respectively, the critical temperature for BEC takes the form \cite{bagnato1987bose}:
\begin{equation}
	T_c^0 = \dfrac{\hbar \omega}{k_B}\left(\dfrac{N}{\zeta(3)}\right)^{1/3},
    \label{eq: ideal gas tc}
\end{equation}
where $\omega=(\omega_x\omega_y\omega_z)^{1/3}$, $\hbar$ is the reduced Planck's constant, $k_B$ is the Boltzmann constant and $\zeta(3)\sim1.2$. For $T<T_c^0$ a BEC is formed and a simple relation between the condensed fraction of atoms $N_{\text{BEC}}/N$ and the temperature of the system can be written as:
\begin{equation}
    \dfrac{N_\text{BEC}}{N} = 1-\left(\dfrac{T}{T_c^0}\right)^3.
    \label{eq: fração condensada}
\end{equation}

The correction to the critical temperature due to interactions, $\delta T_c^\text{int}$, is deduced considering the mean-field interaction potential $V_{\text{int}} = gn(\textbf{r})$ where $g = 4\pi\hbar^2a/m$, $n(\textbf{r})$ is the atomic density and it reads \cite{giorgini1996condensate}:
\begin{equation}
    \dfrac{\delta T_c^\text{int}}{T_c^0}=-1.33 \sqrt{\frac{m\omega}{\hbar}}aN^{\frac{1}{6}}.
    \label{eq: drift 11}
\end{equation}

An additional shift in the critical temperature due to finite size effects is also calculated in \cite{giorgini1996condensate} and given by 
\begin{equation}
    \dfrac{\delta T_c^{\text{size}}}{T_c^0} = -0.73\frac{\Bar{\omega}}{\omega}N^{-1/3},
    \label{eq: drift size}
\end{equation}
where $\Bar{\omega} = \frac{\omega_x + \omega_y + \omega_z}{3}$ is the arithmetic average of the trap frequencies.  

It is expected that in the context of atomic mixtures, the interspecies interaction will cause an additional shift to the critical temperature of a phase transition. For the particular case of Bose-Fermi mixtures, the shift of the BEC critical temperature has been calculated in \cite{albus2002critical} however, the case of bosonic mixtures has remained open, although a recent study estimates it for the simpler situation of a symmetric and thermal mixture \cite{hryhorchak2019large}. In this paper, we cover a broader scenario than only the symmetric mixture, by obtaining an analytical expression for the shift in the critical temperature of the BEC phase transition of a principal species in the presence of a secondary species that could be solved numerically for any bosonic mixture trapped by harmonic potentials fulfilling that $\frac{\omega_{x,1}}{\omega_{x,2}} = \frac{\omega_{y,1}}{\omega_{y,2}} = \frac{\omega_{z,1}}{\omega_{z,2}} = \frac{(\omega_{x,1}\omega_{y,1}\omega_{z,1})^{1/3}}{(\omega_{x,2}\omega_{y,2}\omega_{z,2})^{1/3}} \equiv \frac{\omega_1}{\omega_2}$, which is usually the case for pure magnetic and optical traps~\cite{grimm2000optical, pethick2008bose}.We apply our results to realistic experimental parameters by considering a $^{23}$Na-$^{39}$K mixture \cite{castilho2019compact,gutierrez2021miscibility,schulze2018feshbach} and show that, by changing the atom number ratio between the species, it is possible to obtain a shift that is of the same order as the one resulting from intraspecies interactions and therefore, measurable in current experimental setups.

\section{Single bosonic species}\label{Sec:Theory}


This section summarizes the calculations made in \cite{giorgini1996condensate}, which are later used in an analogous way for the calculation with two species. The starting point to describe a weakly interacting Bose gas is the Gross-Pitaevskii equation, given by 
\begin{equation}
    \left(\dfrac{-\hbar^2}{2m}\nabla^2+V(\textbf{r})+g|\psi(\textbf{r})|^2\right)\psi(\textbf{r})=\mu\psi(\textbf{r}),
    \label{eq:GPE}
\end{equation}
where $\psi(\textbf{r})$ is the wave function, $V(\textbf{r}) = \frac{1}{2}m(\omega_x^2x^2+\omega_y^2y^2+\omega_z^2z^2)$ and $\mu$ is the chemical potential. 

First, using a mean field approximation, the interacting density $n(\textbf{r})$ can be rewritten in terms of the non-interacting density $n^0(\textbf{r})$:
\begin{equation}
    n(\textbf{r}) = n^0(\textbf{r}) - 2gn^0(\textbf{r})\dfrac{\partial n^0}{\partial \mu},
    \label{eq:dens1esp}
\end{equation}
and the density for a non-interacting thermal cloud is well known as  
\begin{equation}
    n^0(\textbf{r}) = (\lambda_T)^{-3}g_{3/2}\left(\exp\left[\dfrac{\mu-V(\textbf{r})}{k_BT}\right]\right),
    \label{eq: density 1}
\end{equation}
where $\lambda_T = \hbar\left(\frac{2\pi}{mk_BT}\right)^{1/2}$ is the thermal wavelength.
Using the normalization condition  $N = \int n(\textbf{r})d\textbf{r}$ and expanding it in first order around $T_c = T_c^0+\delta T_c$ and $\mu = 0$, we find two different shifts:
\begin{equation}
\left\{
	\begin{array}{ll}
\dfrac{\delta T_c^{\text{int}}}{T_c^0}=-\dfrac{2g}{T_c^0}\dfrac{\int d\textbf{r}\frac{\partial n^0}{\partial\mu}[n^0(\textbf{r}=0)-n^0(\textbf{r})]}{\int d\textbf{r}\frac{\partial n^0}{\partial T}}
\vspace{0.1cm}\\
\dfrac{\delta T_c^{\text{size}}}{T_c^0} = -\dfrac{3\hbar\omega}{2T_c^0}\dfrac{\int d\textbf{r}\frac{\partial n^0}{\partial\mu}}{\int d\textbf{r}\frac{\partial n^0}{\partial T}}
	\end{array}
\right.
\label{eq: desviostctamint}
\end{equation}

These equations are easily solved and result in equations \ref{eq: drift 11} and \ref{eq: drift size}, respectively. For the $^{23}$Na parameters found in \cite{gutierrez2021miscibility} with $N=5\times10^6$ atoms, $a=54\,a_0$ and $\omega=2\pi\times\unit[119]{Hz}$, $\bar{\omega}=2\pi\times \unit[120]{Hz}$, the shift due to finite size is around $0.4\%$, while the shift due to interaction is around $2.6\%$. 


\section{Two bosonic species}\label{subsec:TwoSpecies}
For the case of weakly interacting bosonic mixtures, the system is described by a pair of coupled GPEs 
\begin{equation}
\small{
\left\{\begin{matrix}
\left(\dfrac{-\hbar^2}{2m_1}\nabla^2+V_1(\textbf{r})+g_1|\psi_1(\textbf{r})|^2+g_{12}|\psi_2(\textbf{r})|^2\right)\psi_1(\textbf{r})=\mu_1\psi_1(\textbf{r}) \\
\left(\dfrac{-\hbar^2}{2m_2}\nabla^2+V_2(\textbf{r})+g_2|\psi_2(\textbf{r})|^2+g_{12}|\psi_1(\textbf{r})|^2\right)\psi_2(\textbf{r})=\mu_2\psi_2(\textbf{r})
\end{matrix}\right.}
    \label{eq:GPE 2 species}
\end{equation}
where the subindex in each term is related to each one of the two species (e.g., $V_1(\textbf{r})=\frac{1}{2}m_1(\omega_{x,1}^2x^2+\omega_{y,1}^2y^2+\omega_{z,1}^2z^2)$ is the trapping potential of species-1) and $g_{12} = 2\pi\hbar^2a_{12}/m_R$ is the interspecies coupling constant, with $m_R = \frac{m_1m_2}{m_1+m_2}$ the reduced mass. We will consider only the case of repulsive intraspecies interactions, with $g_1,g_2>0$, for which the individual BECs are stable and we will keep $g_{12}^2<g_1g_2$, which is the miscible condition for the homogeneous system, such that we do not expect any phase-separation to occur at the critical point of species-1~\footnote{Even if the miscible condition for a trapped system is not exact, miscibility effects are more relevant when both species exhibit a BEC and, for the range of parameters explored in this paper, we did not see a strong depletion of species-1 due to its interaction with species-2 when running a Hartree-Fock model \cite{pitaevskii2016bose}.}.

Let species-1 to be our main species for which we aim to calculate the shift in $T_{c,1}^0$ due to intra- and interspecies interactions, while species-2 is the secondary species which could be either thermal ($T_{c,1}^0>T_{c,2}^0$) or condensed ($T_{c,1}^0<T_{c,2}^0$). Following the same procedure done in \cite{giorgini1996condensate} for the case of a single species, the interacting density of species-1 here is
\begin{equation}
    n_1(\textbf{r}) =  n_1^0(\textbf{r}) - 2g_{1}n_1^0(\textbf{r})\dfrac{\partial n_1^0}{\partial \mu_1}- g_{12}n_2^0(\textbf{r})\dfrac{\partial n_1^0}{\partial \mu_1},
    \label{eq: densidades mistura}
\end{equation}
where the factor of two is due to exchange effects: species-1 has a two-way interaction, while species-2 interact with species-1 in one-way. Again, using the normalization condition $N_1=\int n_1(\textbf{r})d\textbf{r}$ and expanding in first order around $T_{c,1} = T_{c,1}^0+(\delta T_{c,1})_{11}+(\delta T_{c,1})_{12}$ and $\mu_1=0$, we find that the shift in the critical temperature of species-1 due to interaction effects of species-2 is:
\begin{equation}
    \left(\dfrac{\delta T_{c,1}}{T_{c,1}^0}\right)_{12}=
    - \dfrac{g_{12}}{T_{c,1}^0}\dfrac{\int d\textbf{r}\frac{\partial n_1^0}{\partial\mu_1}[n_2^0(\textbf{r}=0)-n_2^0(\textbf{r})]}{\int d\textbf{r}\frac{\partial n_1^0}{\partial T}}.
\label{eq: delta tc 2esp int}
\end{equation}
In order to solve this equation, one needs to choose the most proper density profile for species-2. This will depend on the regime this atomic cloud is found to be. In the following, we divide the two possible cases considering the second species being above or below its own critical point.


\subsection{Not condensed second species}\label{Subsec:thermal}

Consider first the case for which the system achieves a temperature $T=T_{c,1}^0$, the critical temperature for species-1, for which species-2 is still above its own critical point with $T_{c,2}^0<T_{c,1}^0$. For this case, the density of the second species has the same shape of Eq.~\ref{eq: density 1} and can be written as:
\begin{equation}
    n_2^0(\textbf{r})=(\lambda_{T_{c,1}^0,2})^{-3}g_{3/2}\left(\exp\left[\dfrac{\mu_2-V_{2}(\textbf{r})}{k_BT_{c,1}^0}\right]\right).
    \label{eq: densidade 2 term}
\end{equation}
By replacing $n_2^0(\textbf{r})$ in Eq.~\ref{eq: delta tc 2esp int}, the critical temperature shift of species-1 due to its interaction with species-2 is obtained by evaluating:
\begin{equation}
\begin{aligned}
    \left(\dfrac{\delta T_{c,1}}{T_{c,1}^0}\right)_{12}=&-\dfrac{g_{12}}{3N_1}\dfrac{(\lambda_{T^0_{c,1},1})^{-3}(\lambda_{T^0_{c,1},2})^{-3}}{k_BT_{c,1}^0}\\&\int drg_{1/2}\left(\exp\left[\dfrac{-V_1(\textbf{r})}{k_BT_{c,1}^0}\right]\right)     \\&
    \cdot \left[g_{3/2}\left(\exp\left[\dfrac{\mu_2}{k_BT_{c,1}^0}\right]\right) - \right. \\ & \left. g_{3/2}\left(\exp\left[\dfrac{\mu_2-V_2(\textbf{r})}{k_BT_{c,1}^0}\right]\right)\right].
\end{aligned}
\label{eq: desvio 2 espécie inicio}
\end{equation}
To start solving the previous integral, we will apply the following transformation:
\begin{equation}
\left\{
	\begin{array}{ll}
x'^2=\frac{m_1\omega_{x,1}^2}{2k_BT_{c1}^0}x^2
\\
y'^2=\frac{m_1\omega_{y,1}^2}{2k_BT_{c1}^0}y^2\;\;\Longrightarrow\;\;d\textbf{r} = \left(\frac{2k_BT_{c1}^0}{m_1\omega_1^2}\right)^{3/2}d\textbf{r'},
\\
z'^2=\frac{m_1\omega_{z,1}^2}{2k_BT_{c1}^0}z^2
\end{array}
\right.
\label{eq: substituiçãoint}
\end{equation}
once it simplifies the trapping potentials of each species to $\frac{V_{1}(\textbf{r'})}{k_BT_{c,1}^0}=\alpha\left(x'^2+y'^2+z'^2\right)$ and $\frac{V_2(\textbf{r'})}{k_BT_{c,1}^0}=x'^2+y'^2+z'^2$, with $\alpha=\frac{m_1\omega^2_1}{m_2\omega^2_2}$, considering only  that $\frac{\omega_{x,1}}{\omega_{x,2}} = \frac{\omega_{y,1}}{\omega_{y,2}} = \frac{\omega_{z,1}}{\omega_{z,2}} = \frac{\omega_1}{\omega_2}$. Then, changing to spherical coordinates with $r'=\sqrt{x'^2+y'^2+z'^2}$, $\theta$ and $\phi$, we obtain the shift of $T_{c,1}$ in terms of $\mu_2$

\begin{equation}
\begin{matrix}
    \left(\dfrac{\delta T_{c,1}}{T_{c,1}^0}\right)_{12} =  \dfrac{-g_{12}(\lambda_{T_{c,1}^0,1})^{-3}}{3N_1k_BT_{c,1}^0}\left(\dfrac{k_BT_{c,1}^0}{\hbar \omega_2}\right)^{3}\times \\ \left[\sum^\infty_{n,n'=1}\dfrac{\exp\left({n'\frac{\mu_2}{k_BT_{c,1}^0}}\right)}{n^{1/2}n'^{3/2}}\left(\dfrac{1}{(\alpha n)^{3/2}}-\dfrac{1}{(\alpha n+n')^{3/2}}\right)\right].
    \label{eq: drift termico}
\end{matrix}
\end{equation}
The chemical potential of the species-2 is obtained using the normalization condition $N_2 = \int d\textbf{r} n_2^0(\textbf{r})$, resulting in the relation:
\begin{equation}
        \sum_{n=1}^\infty \dfrac{\exp\left(n\frac{\mu_2}{k_BT_{c,1}^0}\right)}{n^3}=N_2\left(\dfrac{\hbar\omega_2}{ k_BT_{c,1}^0}\right)^{3}.
        \label{eq: mu2 somatorio}
\end{equation}
This normalization condition describes a generic thermal chemical potential that is negative and much larger than the thermal energy $k_BT_{c,1}^0$ when the second species is far from its critical temperature and goes to zero when it is close to $T_{c,2}^0$.

\begin{figure}[h]
    \centering
    \includegraphics[width=0.48\textwidth]{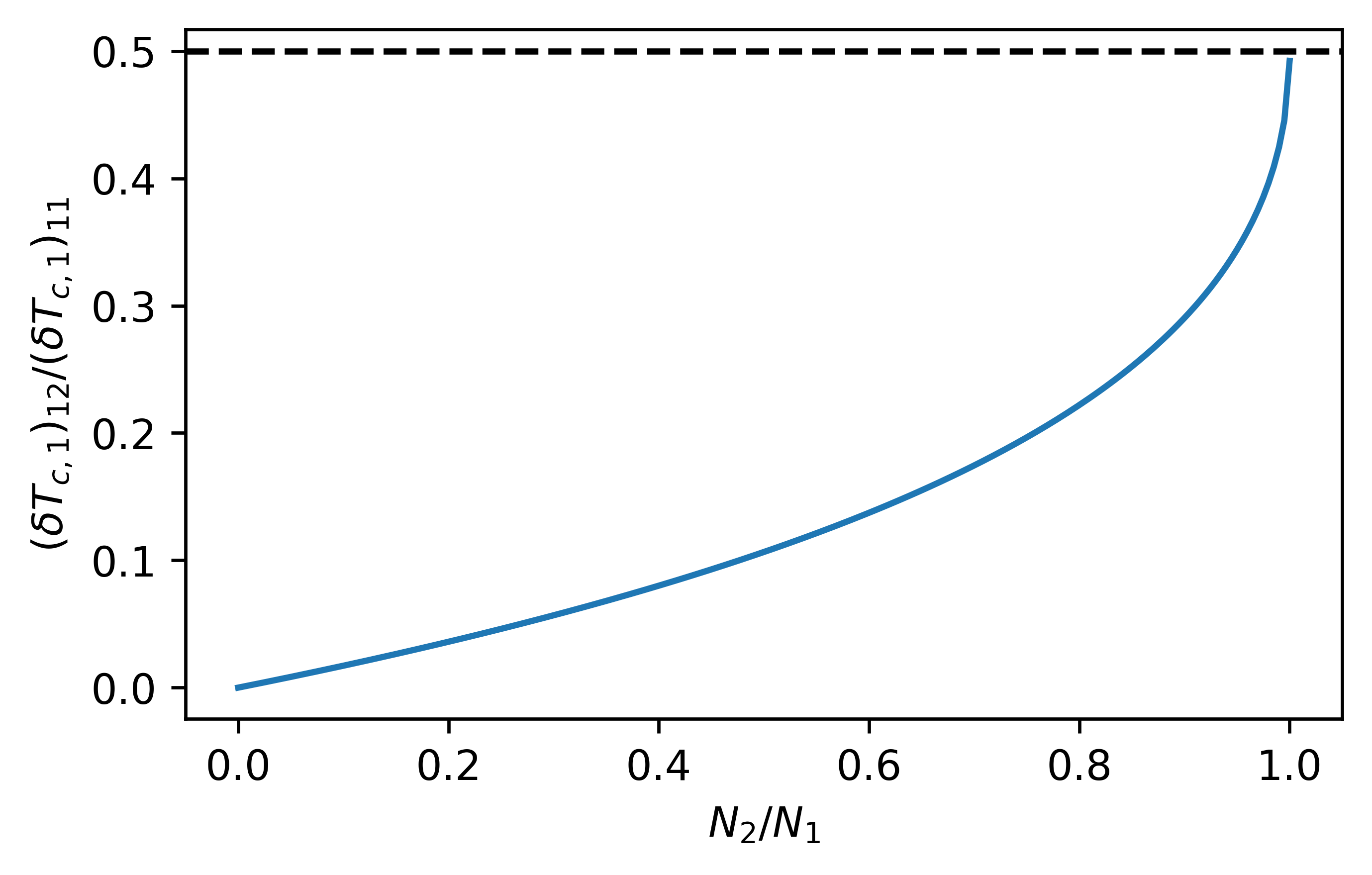}
    \caption{Comparing the shift due to single species and due to the second species for species-1 equal to species-2 when varying $N_2$. The curve reaches exactly 0.5 at $N_2/N_1=1$ as expected thanks to the absence in exchange effects between species-1 and 2 (see the main text).}
    \label{fig: drift term}
\end{figure}

In order to find the shift in the critical temperature of species-1 for a given number of atoms of species-2, $N_2$, one needs to combine Eq.~\ref{eq: drift termico} and~\ref{eq: mu2 somatorio}, solving them numerically, for example, using the bisection method to find the value of $\mu_2$, which is then used to evaluate the shift. To gain some physical intuition on the values we expect for $(\delta T_{c,1})_{12}$ and check our results, we can consider the case of identical species, i.e. $m_1=m_2$, $a_1=a_2=a_{12}$ and $V_1(\textbf{r})=V_2(\textbf{r})$. In this case, we should expect that for $N_{1}=N_{2}$, $T_{c,1}^0=T_{c,2}^0$ and $(\delta T_{c,1})_{12}=0.5(\delta T_{c,1})_{11}$, with $(\delta T_{c,1})_{11}$ given by Eq.~\ref{eq: drift 11}. The factor of half in the previous relation is due to the factor of one in Eq.~\ref{eq: densidades mistura} for interspecies interactions instead of the factor of two for intraspecies interactions related to exchange effects, as already discussed. In Fig.~\ref{fig: drift term}, we show the numerical result of Eq.~\ref{eq: drift termico} normalized by the intraspecies shift when considering identical species and varying the number of atoms of species-2 as $0\leq N_2\leq N_{c,2} =\zeta(3) \left(\frac{k_BT_{c,1}^0}{\hbar\omega_2}\right)^{3}$. The largest limit of $N_2$ represents the critical number for which one saturates the thermal state, meaning that any additional atom would populate the condensed state, and it coincides with the condition of $N_2=N_1$ considered above. The dashed line marks the half condition which is achieved exactly for $N_2=N_1$, as expected.

\subsection{Condensed second species}

Now, let us consider the case where $T_{c,2}^{0}>T_{c,1}^{0}$. In this region, the density of species-2 is composed of a thermal and a condensed fractions. In our simple model, the thermal fraction is represented by equation \ref{eq: densidade 2 term} with $\mu_2=0$ (i.e. related to the critical number), while the condensed fraction is given by the Thomas-Fermi limit \cite{pethick2008bose} -- in the case of a harmonic oscillator, an inverted parabola. The total density can be written as the sum of these two contributions and reads:
\begin{equation}
\begin{matrix}
 n_2^0(\textbf{r}) = (\lambda_{T_{c,1}^0,2})^{-3}g_{3/2}\left(\exp\left[\frac{-V_2(\textbf{r})}{k_BT_{c,1}^0} \right ] \right ) + \\  \text{max}\left[\dfrac{\mu_2}{g_2}\left(1-\dfrac{x^2}{R_{x,2}^2}-\dfrac{y^2}{R_{y,2}^2}-\dfrac{z^2}{R_{z,2}^2} \right ),0\right],
 \end{matrix}
    \label{eq: dens cond}
\end{equation}
where $\mu_2 = \frac{15^{2/5}}{2}\left(N_{\text{BEC},2} a_2\sqrt{\frac{m_2\omega_2}{\hbar}}\right)^{2/5}\hbar\omega_2$ is the chemical potential of the condensed cloud \footnote{Although the chemical potential is zero for the thermal fraction, it is nonzero and positive for the condensed one.}, with $N_{\text{BEC},2}/N_2 = 1-\left(T_{c,1}^0/T_{c,2}^0\right)^3$  the condensed fraction of species-2 and $R_{i,2} = \sqrt{\frac{2\mu_2}{m_2\omega_{i,2}^2}}$ the Thomas-Fermi radius of species-2 in the $\hat{i}$ direction.


Using Eq. \ref{eq: dens cond} in Eq. \ref{eq: delta tc 2esp int} allow us to find the shift in the critical temperature of species-1 when species-2 has some non-vanishing condensed fraction. The contribution of the thermal fraction in this shift continues to be given by Eq. \ref{eq: drift termico} for $\mu_2 = 0$. Now, our task resumes in obtaining the contribution related to the condensed part. For that, it is convenient to normalize the coordinates in the integral in terms of the Thomas-Fermi radius ($i\rightarrow i/R_{i,2}$) together with the use of spherical coordinates given in Eq. \ref{eq: substituiçãoint}. Then, the shift in the critical temperature due to the interaction with the second species, when $T_{c,2}^0<T_{c,1}^0$ is
{\small
\begin{equation}
\begin{aligned}
\left(\frac{\delta T_{c,1}}{T_{c,1}^0}\right)_{12} &= -\dfrac{g_{12}(\lambda_{T_{c,1}^0,1})^{-3}}{3N_1k_BT_{c,1}^0}\left(\dfrac{k_BT_{c,1}^0}{\hbar \omega_2}\right)^{3} \times
\\ & \left[\sum_{n,n'=1}^\infty
\frac{1}{n^{1/2} n'^{3/2}}\left(\frac{1}{(\alpha n)^{3/2}}-\frac{1}{(\alpha n+n')^{3/2}}\right)\right]\\[4pt]
&\quad - \frac{4\pi g_{12}R_xR_yR_z\mu_2 }{3g_2\lambda_{T_{c,1}^0,1}^3N_1k_B T_{c,1}^0}
\sum_{n=1}^\infty \frac{1}{n^{1/2}}
\\ & \int_0^1 \! dr\, r^4 \exp\!\left(-\frac{n\alpha \mu_2}{k_B T_{c,1}^0}\, r^2\right).
\end{aligned}
\label{eq:drift_cond}
\end{equation}
}

\begin{figure}[h]
    \centering
    \includegraphics[width=0.48\textwidth]{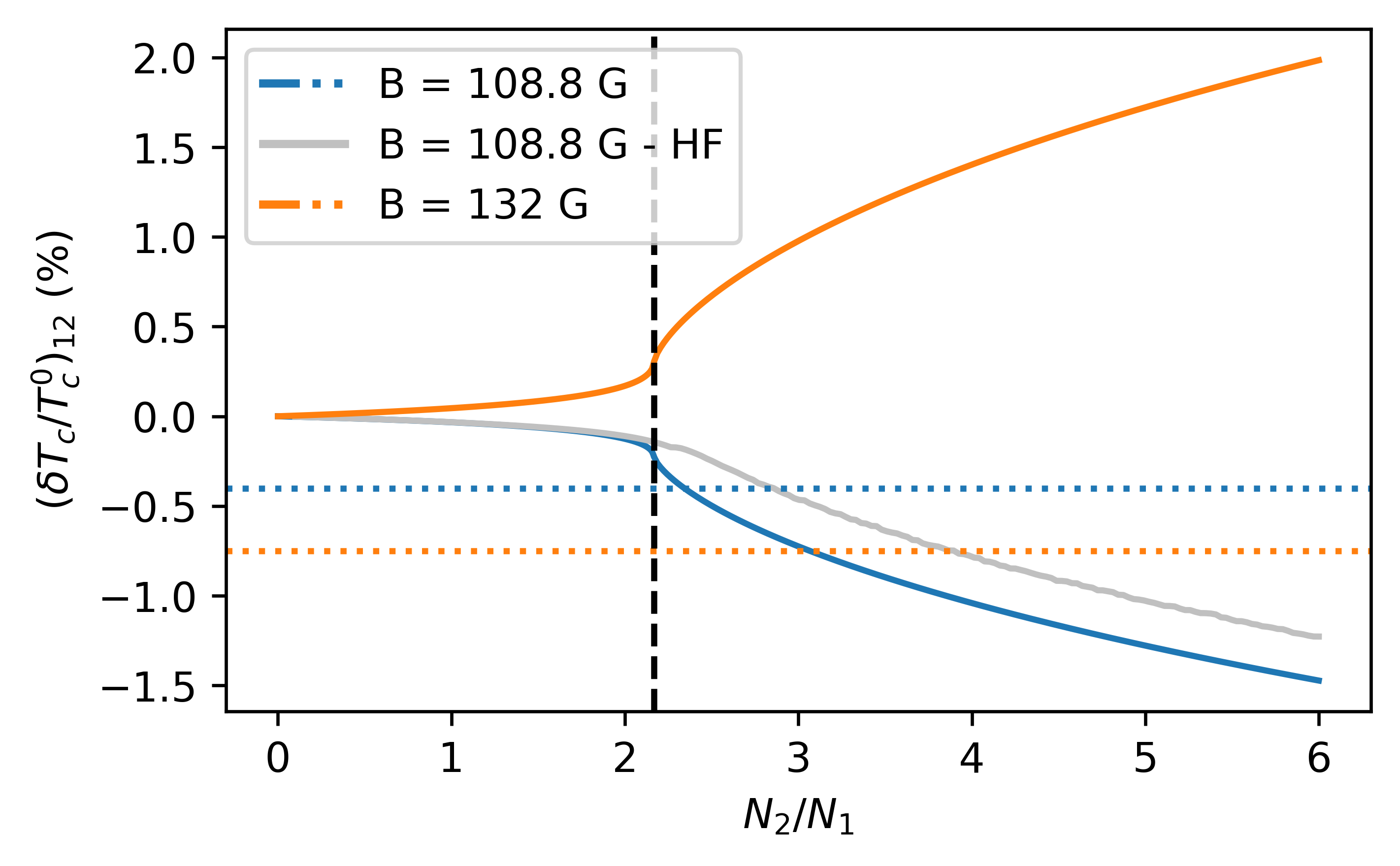}
    \caption{Shift in the critical temperature of species-1 in terms of number of atoms of species-2 for a  $^{23}$Na-$^{39}$K mixture~\cite{castilho2017new} considering $B = 108.8~$Gauss (in blue) and $B = 132~$Gauss (in orange). The lines represents Eq. \ref{eq: drift termico} (before dashed line) and \ref{eq:drift_cond} (after dashed line), while the dotted lines represents Eq. \ref{eq: drift 11} for each magnetic field. The curve in gray represents a full HF simulation of the shift (see supplementary material). The vertical dashed line represents the BEC transition point for species-2, with $T_{c,1}^0=T_{c,2}^0$.} 
    \label{fig: drift cond}
\end{figure}

The previous expression can also be solved numerically considering a specific set of parameters defining the Bose-Bose mixture. Relevant to our purposes, we considered a mixture of sodium and potassium based on parameters found in \cite{gutierrez2021miscibility,schulze2018feshbach}, where potassium is defined as species-1 and sodium as species-2, with related scattering length $a_2=54\,a_0$. The interest in such a mixture is mostly due to the $^{39}$K and $^{39}$K-$^{23}$Na Feshbach resonances existing at moderate magnetic fields ($B<300~$Gauss) that allow to change the sign of the interspecies scattering length as well as the miscibility of the system. In the Brazilian $^{23}$Na-$^{39}$K experiment \cite{castilho2017new}, the atoms are trapped by an optical dipole trap, with frequencies $\omega_1 = 2\pi \times \unit[154]{Hz}$ and $\omega_2 = 2\pi \times \unit[119]{Hz}$ near the critical point for the BEC transition of species-1. In our calculations, we fixed the number of atoms of species-1, $N_1=8\times 10^5~$atoms, and varied the number of atoms of species-2 from zero to $6N_1$ considering two different magnetic fields lying between the first two resonances of $^{39}$K. For $B = 108.8~$Gauss, $a_1 = 7.7a_0$ and $a_{12} = 19.3a_0$ are both positive giving rise to negative shifts $\delta T_{c,1}$. For $B = 132~$Gauss, $a_1 = 14.34a_0$ and $a_{12} = -26a_0$ and the interspecies shift becomes positive making the BEC at species-1 to appear earlier than for the single species case. Fig.~\ref{fig: drift cond} shows the complete curves, Eqs.~\ref{eq: drift termico} and~\ref{eq:drift_cond}, for the shift in the critical temperature of species-1 due to the interaction with species-2 (blue line for $B=108.8~$Gauss and orange line for $B = 132~$Gauss) in comparison with the intraspecies shift due to Eq.~\ref{eq: drift 11} for each case (dotted lines). The vertical dashed line indicates the condition for which $T_{c,2}^0=T_{c,1}^0$ representing the critical point of species-2. We compared our results with a full self-consistent Hartree-Fock model~\cite{pitaevskii2016bose} for which the thermal cloud of species-2 suffers a central depletion as a result of the interaction with its BEC fraction~\footnote{See Supplementary Material.}. The correction do the data of Fig.~\ref{fig: drift cond} is small at the critical point and it tends to the value of the shift due to the thermal part of Eq.~\ref{eq:drift_cond} as the BEC increases and the thermal cloud gets more and more depleted. So, for the most interesting case where species-2 exhibit a large condensed fraction, our simple equations may be corrected a posterior.

Then, by abruptly changing $N_2$ in the vicinity of the critical point for species-1 while still in the thermal regime, let it say, by simply removing the sodium atoms when $a_{12}$ is positive, the remaining atomic cloud will instantaneously cross its critical point in such a way that the control parameter is no longer the temperature of the system. The possibility of studying the phase transition induced by a secondary species can be an interesting outcome of the results presented in Fig.~\ref{fig: drift cond}, where the magnitude and direction of the shift could be controlled by the strength and sign of the interspecies interaction.

\section{Conclusions}


In ultracold gases, understanding how the critical temperature varies with the interaction strength, whether in single-species systems or mixtures, is extremely useful for designing experiments and for investigations near the phase-transition's point. In this manuscript, we discussed the two possible scenarios when considering a Bose-Bose mixture, with the secondary species above or below its own critical temperature, and obtained expressions for the shift in the critical temperature of a principal species in each case which could be easily solved numerically. The expressions obtained here can be extended to any Bose-Bose mixture in all trapping geometries. For the single-species systems, the shift at $T_c^0$ has been measured in different scenarios using standard imaging techniques and thermometry methods \cite{gerbier2004critical,smith2011effects}. The possibility of having interspecies shifts of the same order of magnitude or larger than the intraspecies one, as demonstrated in Fig.~\ref{fig: drift cond}, allows it to be measured in current running experiments without the need of specific and more precise techniques.

The ability to manipulate the transition temperature by controlling the atom number ratio and/or the interspecies scattering length offers the possibility of inducing a phase transition by changing any of these two parameters and creating a complex phase-diagram. By varying the concentration of each species, going from $0\%$ to a $100\%$ for species-1 or -2, it is possible to identify regions where the system contains a single condensate, two condensates, or simply a mixture of thermal gases. This type of phase diagram, reminiscent of binary alloys, may reveal a rich structure of domains~\cite{okamoto1990binary}, which remains largely unexplored in the context of ultracold atomic gases. One may view a mixture of condensates as a binary superfluid system, where the balance between inter- and intra-species interactions determines domains of varying miscibility, giving rise to a diverse landscape of textures and structural features within the system.

\textit{Acknowledgments.} We thank K.M. Farias and E.G.I. Salcedo for early discussions on the development of this project. This work was supported by the S\~ao Paulo Research Foundation (FAPESP) under the grants 2013/07276-1, 2021/09920-1 and 2024/04219-1, and by the Coordination of Superior Level Staff Improvement (CAPES) under the grant 88887.965507/2024-00.

\textit{Author contributions.} P.M.G. provided the analytical and numerical results; All authors participated in the discussions and in the writing of the manuscript; P.C.M.C supervised the project.

\nocite{*}

\bibliographystyle{apsrev4-2}
\bibliography{ref}

\newpage
\clearpage

\pagebreak
\twocolumngrid
\begin{center}
\textbf{\large Supplemental Material: \TITLE}
\end{center}
\setcounter{equation}{0}
\setcounter{figure}{0}
\setcounter{table}{0}
\makeatletter
\renewcommand{\theequation}{S\arabic{equation}}
\renewcommand{\thefigure}{S\arabic{figure}}
\renewcommand{\bibnumfmt}[1]{[S#1]}

\section{Comparison with Hartree-Fock model}

A more appropriate model when dealing with weakly-interacting Bose gases at finite temperature is the Hartree-Fock approximation which will consider the effect of interactions of both thermal and condensate clouds~\cite{pitaevskii2016bose}. The drawback in choosing this approach, is that one should self-consistently solve the following set of equations:
\begin{equation}\label{eq:HF-BEC}
    \left(\dfrac{-\hbar^2}{2m}\nabla^2+V(\textbf{r})+g[n_0(\textbf{r}) + 2n_{th}(\textbf{r})]\right)\psi_0(\textbf{r})=\mu\psi_0(\textbf{r}),
\end{equation}

\begin{equation}\label{eq:HF-thermal}
    \left(\dfrac{-\hbar^2}{2m}\nabla^2+V(\textbf{r})+2gn(\textbf{r})\right)\phi_i(\textbf{r})=\epsilon_i\phi_i(\textbf{r}),
\end{equation}

\begin{equation}\label{eq:HF-Normalization}
    N = \int d\textbf{r}n(\textbf{r}) = \int d\textbf{r}[n_0(\textbf{r}) + n_{th}(\textbf{r})],
\end{equation}
where $n(\textbf{r})$ is the total atomic density, $n_0(\textbf{r}) = \vert\psi_0(\textbf{r})\vert^2$ and $\psi_0(\textbf{r})$ are the condensate density and the condensate wavefunction, respectively, $\phi_i(\textbf{r})$ is the wavefunction of state-$i$ with energy $\epsilon_i$ and $i>0$ and $n_{th}(\textbf{r}) = \sum_{i\neq0}n_i\vert\phi_i(\textbf{r})\vert^2$ is the thermal cloud density, with $n_i = \left[\text{exp}(\beta[\epsilon_i-\mu])-1\right]^{-1}$ being the Bose-Einstein distribution function. Solving these equations usually demand a complex code and some computational power.

A useful approximation considers the BEC to be given in the Thomas-Fermi limit. Therefore, by neglecting the kinetic term in Eq.~\ref{eq:HF-BEC}, one can easily write the condensed and thermal cloud densities as:
\begin{equation}\label{eq:n0_density_HF}
    n_0(\textbf{r})=(\mu_0 - V(\textbf{r}) - 2gn_{th}(\textbf{r}))/g,
\end{equation}
\begin{equation}\label{eq:nTh_density_Veff}
    n_{th}(\textbf{r}) = \frac{1}{\lambda_T^3}g_{3/2}\left(e^{-(V_{eff}-\mu)/k_BT}\right),
\end{equation}
where $g_{3/2}(z)$ is the Bose function or Polylogarithm function $\text{Li}_s(z)=\sum_{k=1}^\infty \frac{z^k}{k^s}$ and $V_{eff} = V(\textbf{r}) + 2gn(\textbf{r})$. The chemical potential $\mu$ is obtained from the normalization condition condition of Eq.~\ref{eq:HF-Normalization} and one can solve the previous two equations for different values of $T/T_c^0$, with $T_c^0$ being the ideal critical temperature.

\begin{figure}[h]
    \centering
    \includegraphics[width=0.48\textwidth]{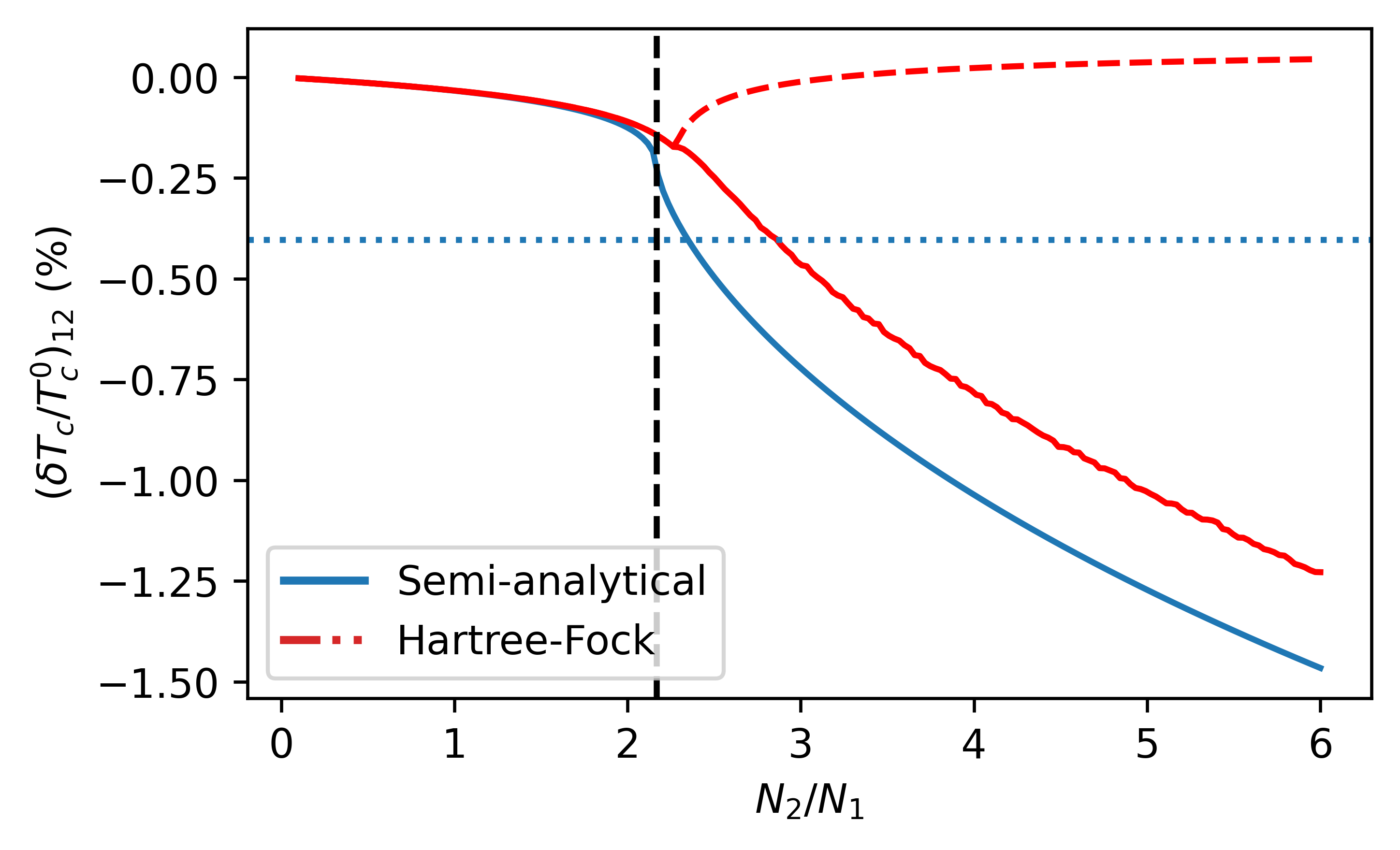}
    \caption{Comparison of the $\delta T_{c,1}/T_{c,1}^0$ obtained from Eq.~\ref{eq:drift_cond} of the main text (in blue) and from the HF model (in red) for $B = 108.8~$Gauss. The dashed red line indicates the contribution due to the thermal cloud in the HF that decreases as a result of its depletion due to the interaction with the BEC. The dotted blue line and the dashed vertical line are the same as in Fig.~\ref{fig: drift cond} of the main text.}
    \label{fig: drift cond - SupMat}
\end{figure}

In order to compare our semi-analytical results with the Hartree-Fock model, we solved Eqs.~\eqref{eq:n0_density_HF} and~\eqref{eq:nTh_density_Veff} for a grid $100\times 100\times 100$ points going from $-50~\mu$m to $+50~\mu$m in each direction which was sufficient to capture the whole cloud without too much loss of resolution. In Fig.~\ref{fig: drift cond - SupMat}, we display the same curves of $\delta T_{c,1}/T_{c,1}^0$ from Fig.~\ref{fig: drift cond} of the main text for $B=108.8~$Gauss (in blue) together with the HF results (in red). Here, we explicitly show the contribution of the thermal cloud above the critical point of species-2 (in dashed), which strongly decreases in the Hartree-Fock model thanks to the depletion in its density distribution resulting from the repulsion of the BEC.

\begin{figure}[h]
    \centering
    \includegraphics[width=0.48\textwidth]{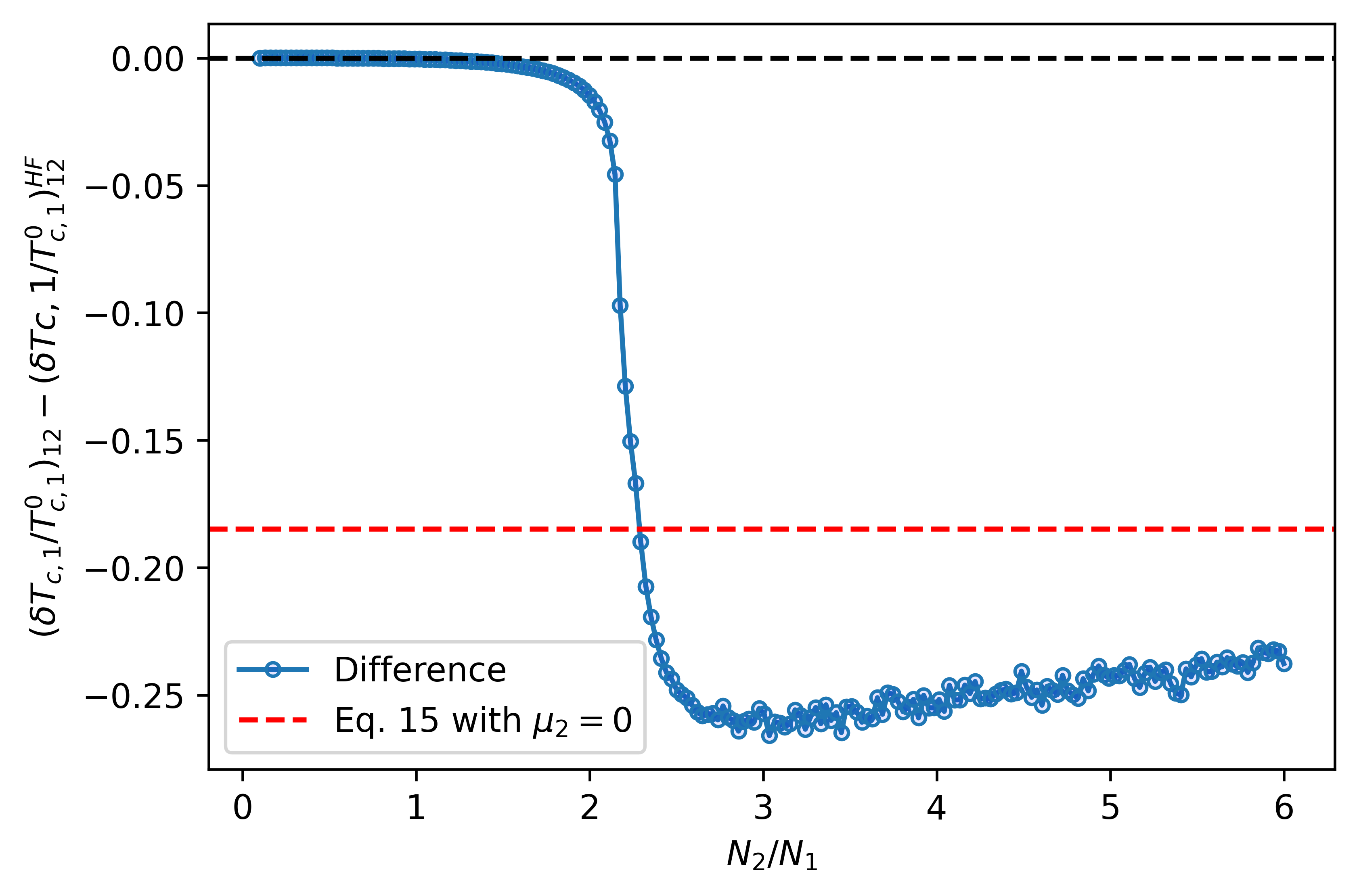}
    \caption{Difference between the shifts of the critical temperature of $^{39}$K in the presence of $^{23}$Na for $B = 108.8~$Gauss obtained from Eq.~\ref{eq:drift_cond} of the main text and from the HF model. The dashed red line represents the shift resulting from Eq.~\ref{eq: drift termico} of the main text, with $\mu_2=0$. Increasing the BEC fraction of species-2 makes the difference to approach the red dashed line.}
    \label{fig: drift cond - SupMat_diff}
\end{figure}

The two curves of Fig.~\ref{fig: drift cond - SupMat} are clearly different, however, one can quantify it by plotting the difference between them, as in Fig.~\ref{fig: drift cond - SupMat_diff} . Here, we see that the difference between the shifts is null for small $N_2/N_1$ ratios, indicating that both models agree when only the thermal cloud is present and we are still far from the critical point of the transition. When approaching the critical point, the difference between the curves strongly changes and for large values of $N_2/N1$ it approaches the value of the shift for an ideal thermal cloud at the critical point (Eq.~\ref{eq: density 1} of the main text with $\mu=0$) given by the dashed black line. This is an important conclusion since, for the most interesting case where a large condensed fraction is present resulting in a larger shift, the simple semi-analytical equation developed in the main text can be corrected a posterior by subtracting the shift of the dashed line without the need of a self-consistent Hartree-Fock simulation.

\end{document}